\documentclass{PoS}

\usepackage{graphicx}
\usepackage{epstopdf}
\usepackage{bm}
\usepackage{epsfig}
\usepackage{graphics}
\usepackage{xspace}
\usepackage{amsmath}
\usepackage{xcolor}

\def\OMIT#1{}

\newcommand{\df}{\rm d}

\newcommand{\bea}{\begin{eqnarray}}
\newcommand{\eea}{\end{eqnarray}}

\newcommand{\gsim}{\mathrel{\rlap{\lower4pt\hbox{\hskip1pt$\sim$}}\raise1pt\hbox{$>$}}}

\newcommand{\Pythia}{\textsc{Pythia}\xspace}

\newcommand{\be}{\begin{equation}}
\newcommand{\ee}{\end{equation}}

\title{Monte Carlo Top Quark Mass Calibration}

\ShortTitle{Monte Carlo Top Quark Mass Calibration}
   
\author{Bahman Dehnadi,$^a$ Andr\'e H. Hoang,$^{bc}$ Vicent Mateu,$^{de}$ \mbox{\speaker{Moritz Preisser}$^{b}$} and Iain W. Stewart$^{f}$\\
\llap{$^a$}Universit\"at Siegen, Department Physik, Walter-Flex-Stra\ss{}e 3, D-57068 Siegen, Germany\\
\llap{$^b$}University of Vienna, Faculty of Physics, Boltzmanngasse 5, A-1090 Wien, Austria\\
\llap{$^c$}Erwin Schr\"odinger International Institute for Mathematical Physics, University of Vienna,\\ Boltzmanngasse 9, A-1090 Wien, Austria\\
\llap{$^d$}Departamento de F\'isica Fundamental and IUFFyM, Universidad de Salamanca, E-37008
Salamanca, Spain\\
\llap{$^e$}Instituto de F\'isica Te\'orica UAM-CSIC, E-28049 Madrid, Spain\\
\llap{$^f$}Center for Theoretical Physics, Massachusetts Institute of Technology,\\ Cambridge, MA 02139, USA\\
E-mail: \email{dehnadi@physik.uni-siegen.de}, \email{andre.hoang@univie.ac.at}, \email{vmateu@usal.es}, \email{moritz.preisser@univie.ac.at}, \email{iains@mit.edu}}

\abstract{The most precise top quark mass measurements use kinematic reconstruction methods, determining the top mass parameter of a Monte Carlo event
generator, $m_t^{\rm MC}$. Due to the complicated interplay of hadronization and parton shower dynamics in Monte Carlo event generators relevant for kinematic reconstruction, relating $m_t^{\rm MC}$ to field theory masses is a non-trivial task. In this talk
we report on a calibration procedure to determine this relation using hadron level QCD predictions for  2-Jettiness in $e^+e^-$ annihilation, an observable which has kinematic top mass sensitivity and a close relation to the invariant mass of the particles coming from the top decay. The theoretical ingredients of the QCD prediction are reviewed.
Fitting $e^+e^-$ 2-Jettiness calculations at NLL/NNLL order to \Pythia~8.205, we find that
$m_t^{\rm MC}$ agrees with the MSR mass $m_{t,1\,{\rm GeV}}^{\rm MSR}$ within uncertainties. At NNLL we find $m_t^{\rm MC} = m_{t,1\,{\rm GeV}}^{\rm MSR} + (0.18 \pm 0.22)\,{\rm GeV}$. $m_t^{\rm MC}$ can differ from the pole mass $m_t^{\rm pole}$ by up to $600\,{\rm MeV}$, and using the pole mass generally leads to larger uncertainties. At NNLL we find $m_t^{\rm MC} = m_t^{\rm pole} + (0.57 \pm 0.28)\,{\rm GeV}$ as the fit result. In contrast, converting $m_{t,1\,{\rm GeV}}^{\rm MSR}$ obtained at NNLL to the pole mass gives a result for $m_t^{\rm pole}$ that is substantially larger and incompatible with the fit result. We also explain some theoretical aspects relevant for employing the C-parameter as an alternative calibration observable.
\\[.5cm]
SI-HEP-2018-10, IFT-UAM/CSIC-18-024, MIT-CTP/4990, UWTHPH 2018-11}

\FullConference{
13th International Symposium on Radiative Corrections (Applications of Quantum Field Theory to Phenomenology)\\
                 25-29 September, 2017 \\
                 St. Gilgen, Austria}

\begin{document}

\section{ Introduction}
The most precise measurements of the top quark mass 
are based on direct reconstruction methods exploiting its kinematic properties and 
have reached uncertainties of about $0.5$\,GeV~\cite{Tevatron:2014cka,Khachatryan:2015hba,Aaboud:2016igd}. They are based on multivariate fits that use a maximum amount of information from the top decay final states.
This includes template and matrix element fits for distributions such as the measured invariant mass. Since these observables are highly differential and depend on experimental cuts and details of the jet dynamics, multipurpose Monte Carlo (MC) event generators are employed in these analyses, and the measured mass is the top mass parameter $m_t^{\rm MC}$ contained in the particular MC event generator. Clearly, the interpretation of $m_t^{\rm MC}$ from the field theoretic point of view is influenced by the interplay of both perturbative and non-perturbative QCD effects and -- because MC generators provide only approximate descriptions -- may also depend in part on the MC tuning and the set of observables used in the analyses. 
In the direct reconstruction analyses referred to above the systematic uncertainties from MC modeling are a dominant part of the uncertainty budget, but they do not address in any way how $m_t^{\rm MC}$ is related to a mass parameter defined precisely in quantum field theory that can be globally used for higher order theoretical predictions. The relation is nontrivial because it requires an understanding of the interplay between the partonic components of the MC generator (hard
matrix elements and parton shower) and the hadronization model. Furthermore it may be affected by common basic approximations made in the construction of the MC machinery. One can also say that -- at the level of precision achieved for top mass measurements in direct reconstruction -- MC generators should be considered as models whose partonic components and hadronization models are, through the tuning procedure, capable
of describing  experimental data to a precision that is higher than that of their partonic input.

In the past $m_t^{\rm MC}$ has frequently simply been identified with the pole mass, which is, however not defined beyond perturbation theory. This identification may not immediately look incompatible with parton-shower implementations for massive quarks, but a direct identification is disfavored because the parton-shower does not account for perturbative corrections from momenta below the MC shower cutoff $\Lambda_c\sim 1\,{\rm GeV}$ and as a consequence, $m_t^{\rm MC}$ may be also sensitive to non-perturbative effects from momenta below $\Lambda_c$. Also, the pole mass has an ${\cal O}(\Lambda_{\rm QCD})$ renormalon ambiguity, while $m_t^{\rm MC}$ does not, since information from perturbative QCD is not employed below $\Lambda_c$. It has been argued~\cite{Hoang:2008xm,Hoang:2014oea} that $m_t^{\rm MC}$ has a much closer (and perturbatively  more stable) relation to the MSR mass~\cite{Hoang:2008yj} $m_t^{\rm MSR}(R\approx \Lambda_c)$, where the scale $R$ defining this scheme is close to $\Lambda_c$.

For a given MC generator (which in this context also implies that one considers a given tune), $m_t^{\rm MC}$ can be calibrated with respect to a field theory mass scheme through a fit of MC predictions to {\it hadron level} QCD computations for observables closely related to the distributions that enter the
experimental reconstruction analyses. In Ref.~\cite{Butenschoen:2016lpz} a precise quantitative study was provided on the interpretation of $m_t^{\rm MC}$ in terms
of the MSR and pole mass schemes based on a hadron level prediction for the 2-Jettiness variable $\tau_2$~\cite{Stewart:2010tn} for the production of a boosted top-antitop quark pair in $e^+e^-$ annihilation. To be definite $\tau_2$ is defined as:
\begin{equation}
\label{tau2def}
\tau_2= 1-\max_{{\vec n}_t}\frac{\sum_i|{\vec n}_t\cdot \vec p_i|}{Q}\,,
\end{equation}
where the sum is over the 3-momenta of all final state particles, the maximum defines the thrust axis $\vec n_{\rm t}$ and $Q$ is the
center of mass energy. In Refs.~\cite{Fleming:2007qr,Fleming:2007xt} a factorization theorem has been proven for boosted top quarks,
yielding hadron level predictions for $\tau_2$.
The $\tau_2$
distribution has a distinguished peak very
sensitive to the top mass, and is a delta function at $\tau_2^{\rm min}(m_t)=1\,-\,\sqrt{1-4 m_t^2/Q^2}$ at tree level. For $Q$ sufficiently larger than $m_t$ the peak region is dominated by dijet events where the top quarks decay inside narrow back-to-back cones, and there $\tau_2$ is directly related to the sum of the squared invariant masses $M^2_{a,b}$ in the two hemispheres defined
by the thrust axis $\vec n_t$, $(\tau_2)_{\rm peak }\approx (M_a^2+M_b^2)/Q^2$~\cite{Fleming:2007qr,Fleming:2007xt}. Therefore $\tau_2$ in the peak region is an observable with direct kinematic top mass sensitivity, just like those that enter the top quark mass reconstruction methods, and the results of the calibration study should provide information relevant for the interpretation of the direct reconstruction measurements.

\section{ 2-Jettiness Distribution}
The $\tau_2$ distribution in the peak region for boosted top quarks has the basic form
\begin{equation}\label{eq:factheo0}
\frac{\df\sigma}{\df\tau_2} \!= \!\!\int\! \df k\, \bigg(\! \frac{\df\hat\sigma_{\rm s}}{\df\tau_2} +
\frac{\df\hat\sigma_{\rm ns}}{\df\tau_2} \!\bigg)\!\!\bigg(\!\tau_2-\frac{k}{Q}\!\bigg)F_{\!\tau_2}\!(k)\!
\bigg[1+{\cal O} \bigg(\!\frac{\Lambda_{\rm QCD}}{Q}, \frac{\Gamma_t}{m_t}\! \bigg)\bigg],
\end{equation}
where $\df\hat\sigma_{\rm s}/\df\tau_2$ contains the singular partonic QCD corrections
$\alpha_s^j\,[\,\ln^k(\tau_2-\tau_2^{\rm min})/(\tau_2\,-\,\tau_2^{\rm min})\,]_+$ and $\alpha_s^j\,\delta(\tau_2 \,-\, \tau_2^{\rm min})$
in the dijet limit and $\df\hat\sigma_{\rm ns}/\df\tau_2$ stands for the remaining partonic nonsingular QCD corrections. The shape function
$F_{\tau_2}$ describes non-perturbative effects from wide-angle soft gluon radiation~\cite{Korchemsky:1999kt}. The singular partonic
contribution obeys the factorization theorem
\begin{align}\label{eq:factheo}
&\frac{\df\hat\sigma_{\rm s}}{\df\tau_2} = \;Q\,H_Q^{(6)}(Q,\mu_Q)\,U_{H_Q}^{(6)}(Q,\mu_Q,\mu_m)\,H_m^{(6)}(Q,m_t,\mu_m)\, U_{H_m}^{(5)}\Big(\frac{Q}{m_t},\mu_m,\mu_B\Big)
  \\
&\times\!{\int} {\df s}{\int}{\df k}\,
J_{B,\tau_2}^{(5)}\Big(\frac{s}{m_t},\mu_B,\Gamma_t,\delta m_t\Big) 
\, U_S^{(5)}(k,\mu_B,\mu_S)\,\hat S_{\tau_2}^{(5)}\Big(Q[\tau_2-\tau_2^{\rm min}(m_t)] - \frac{s}{Q} - k,\mu_S\Big), 
  \nonumber
\end{align}
which is based on Soft-Collinear-Effective Theory~\cite{Bauer:2000ew, Bauer:2000yr, Bauer:2001ct, Bauer:2001yt} and separates the contributions
from the hard interactions in the hard functions $H_Q$ and $H_m$, the jet function $J_{B,\tau_2}$, and the soft cross-talk between the top
and antitop jets in the partonic soft function $\hat S_{\tau}$. The jet function $J_{B,\tau_2}$ is derived in boosted HQET~\cite{Fleming:2007qr}
since the collinear top jet invariant mass in the peak region is very close to the nominal top quark mass. It includes the collinear dynamics of the
decaying top quarks and leading top finite-width effects. The various evolution factors $U_X$ sum large logarithms.

Results for $\df\hat\sigma_s/\df\tau_2$ with next-to-leading logarithmic resummation~$+\,{\cal O}(\alpha_s)$~singular corrections
(NLL\,+\,NLO) can be found in Ref.~\cite{Fleming:2007xt}, with the addition of the virtual top quark contribution and rapidity logarithms
in $H_m$ and the corresponding evolution factor $U_{H_m}$ both from Ref.~\cite{Hoang:2015vua}. The N$^2$LL evolution in $U_{H_Q}$ and $U_S$ is known from the massless quark case, and is consistent with the direct ${\cal O}(\alpha_s^2)$ calculation of the $J_{B,\tau_2}$ anomalous dimension~\cite{Jain:2008gb}.
We implemented all the N$^2$LL order ingredients for the proper treatment of the flavor number dependence
[superscript (6) for including top as dynamic quark versus superscript (5) for excluding the top] in the RG
evolution~\cite{Gritschacher:2013pha, Pietrulewicz:2014qza}. We also include the ${\cal O}(\alpha_s)$ nonsingular corrections
$\df\hat\sigma_{\rm ns}/\df\tau_2$~\cite{BahmanPhD}.

For the shape function $F_{\tau_2}$ we use the convergent basis functions introduced in Ref.~\cite{Ligeti:2008ac} truncated to $4$
elements (where the 4-th element is already numerically irrelevant), which determine the moments of the shape function $\Omega_i$~\cite{Abbate:2010xh,Abbate:2012jh}.
The leading power correction $\Omega_1$ is defined in the \mbox{R-gap} scheme such that it cancels an $\mathcal{O}(\Lambda_{\rm QCD})$ renormalon
present in $\hat S_{\tau_2}$~\cite{Hoang:2007vb}.  
This renders  $\Omega_1$ dependent on the subtraction scale $R_S$, and
we quote results for $\Omega_1$ at
the reference scales $R_S=2$\,GeV.  Equation~(\ref{eq:factheo}) is written in terms of a generic mass scheme $m_t$, with $\delta m_t = m_t^{\rm pole} - m_t$ in $J_{B,\tau_2}^{(5)}$
controlling the dominant sensitivity to the mass scheme. In the pole mass scheme $\delta m_t=0$.  Using renormalon-free schemes, the
$\overline{\rm MS}$ mass with $\delta m_t\propto m_t$ is appropriate for the hard functions. In the jet function $J_{B,\tau_2}^{(5)}$ one
has to adopt a scheme such as MSR~\cite{Hoang:2008yj} with 
$\delta m_t\sim R\sim \Gamma_t$ to avoid upsetting the power
counting in the peak region.
The evolution of the MSR mass with $R$ and of $\Omega_1$ with $R_S$ is described by R-evolution~\cite{Hoang:2008yj,Hoang:2009yr}.
To sum large logarithms we use $\tau_2$-dependent scales $\mu_i(\tau_2)$ and $R_i(\tau_2)$,
which can be expressed in terms of $9$ parameters. These parameters are varied to estimate
perturbative uncertainties.

For a given center of mass energy $Q$, the key parameters that enter the QCD factorization predictions for the $\tau_2$ distribution are
the top mass $m_t$, the top width $\Gamma_t$, the hadronic parameters $\Omega_i$, and the strong coupling $\alpha_s(m_Z)$.  We will
consider fits both in the pole and the MSR mass schemes. The results in the MSR scheme are quoted in terms of
$m_t^{\rm MSR}(1\,\mbox{GeV})$ following~\cite{Hoang:2008xm,Hoang:2014oea}.

\section{ Fit Procedure}
For a given $m_t^{\rm MC}$ we produce MC datasets for $\df\sigma/\df\tau_2$ in the peak region for various $Q$ values. For a given profile
and value of $\alpha_s(m_Z)$ we fit the parameters $m_t$ and $\Omega_i$ of the hadron level QCD predictions to this MC dataset.
For each $Q$ value the distribution is normalized over the fit range, and multiple $Q$s are needed simultaneously to break degeneracies concerning $m_t$ and the soft function moments $\Omega_i$.
We construct the $\chi^2$-function using the statistical uncertainties
in the MC datasets. We do the fit by first, for a given value of $m_t$, minimizing $\chi^2$ with respect to the $\Omega_i$ parameters. The
resulting marginalized $\chi^2$ is then minimized with respect to $m_t$ used in the QCD predictions. Uncertainties obtained for the QCD 
parameters from this $\chi^2$ simply reflect the MC statistical uncertainties used to construct the $\chi^2$.  
To estimate the perturbative uncertainty in the QCD predictions we take $500$ random points in the profile-function parameter space and
perform a fit for each of them. The $500$ sets of best-fit values provide an ensemble from which we remove the upper and lower $1.5\%$ in
the mass values to eliminate potential numerical outliers. From this we then determine central values by averaging the largest
and smallest values and perturbative uncertainties from half the covered interval.

To illustrate the calibration procedure we use \Pythia~8.205~\cite{Sjostrand:2006za,Sjostrand:2014zea} with the $e^+e^-$ default tune~7
(the Monash 2013 tune~\cite{Skands:2014pea} for which $\Lambda_{c} = 0.5$\,GeV) for top mass parameter values $m_t^{\rm MC}=170$, $171$,
$172$, $173$, $174$ and $175$\,GeV. We use a fixed top quark width $\Gamma_t=1.4\,{\rm GeV}$ which is independent of $m_t^{\rm MC}$.
No other changes are made to the
default settings. To minimize statistical uncertainties we generate each distribution with $10^7$ events. We have carried out fits for
the following seven \mbox{Q sets} (in GeV units): $(600, 1000, 1400)$, $(700, 1000, 1400)$, $(800, 1000, 1400)$, $(600$ -- $900)$,
$(600$ -- $1400)$, $(700$ -- $1000)$ and $(700$ -- $1400)$, where the ranges refer to steps of $100$. For each one of these sets we have
considered three ranges of $\tau_2$ in the peak region: $(60\%, 80\%)$, $(70\%, 80\%)$ and $(80\%, 80\%)$, where $(x\%, y\%)$ means that
we include regions of the spectra whose $\tau_2 < \tau_2^{\rm peak}$ having cross-section values larger than $x\%$ of the peak height,
and $\tau_2 > \tau_2^{\rm peak}$ with cross sections larger than $y\%$ of the peak height, where $\tau_2^{\rm peak}$ is the peak position.
This makes a total of $21$ fit settings each of which gives central values and scale uncertainties for the top mass and the $\Omega_i$.

\begin{figure}[t!]
\begin{center}
\includegraphics[width=0.9\textwidth]{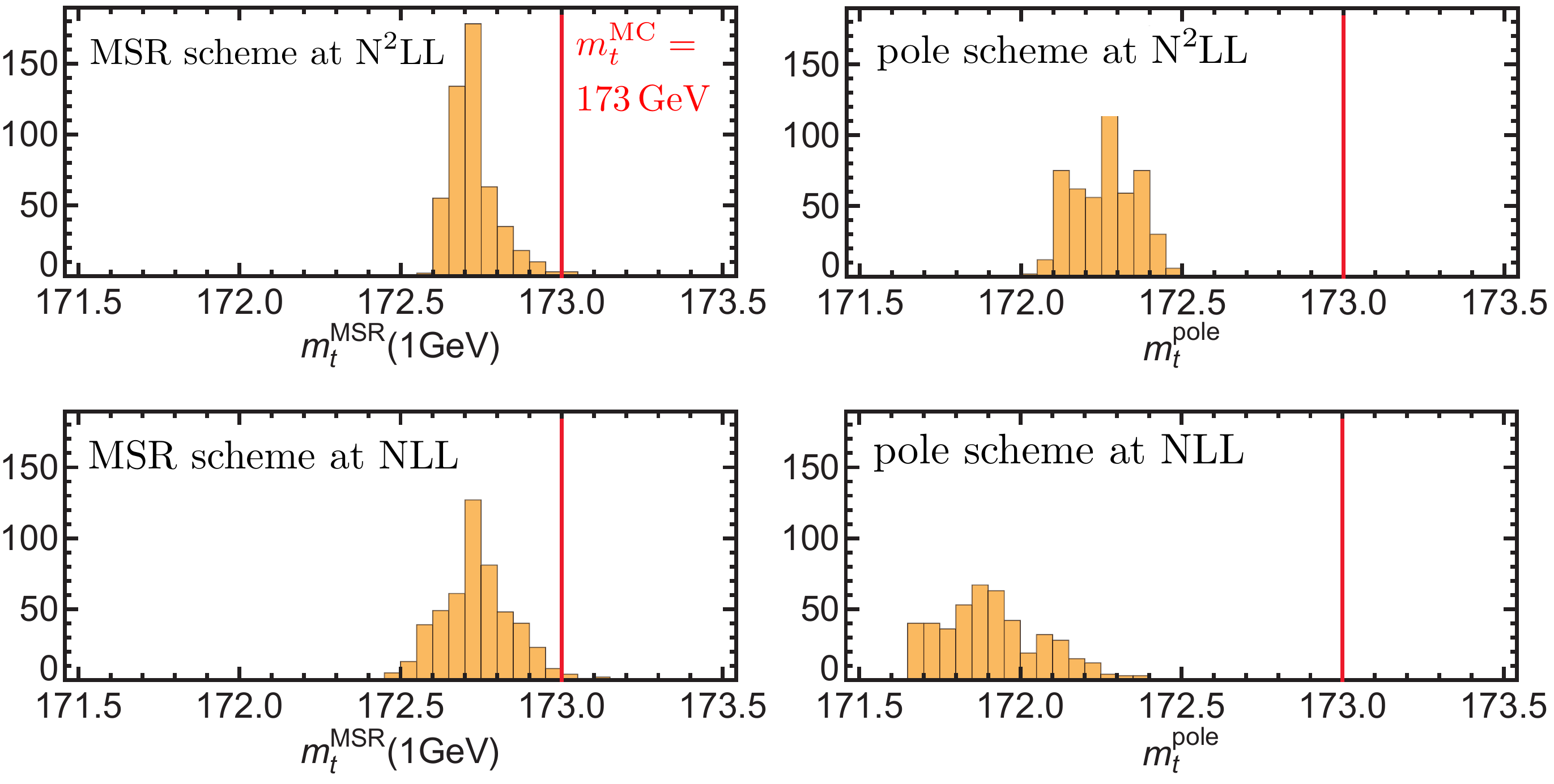}
\caption{\label{fig:histograms} 
Distribution of best-fit mass values from the scan over parameters describing perturbative uncertainties. Results are shown for cross
sections employing the MSR mass $m_t^{\rm MSR}(1\,{\rm GeV})$ (left) and the pole mass $m_t^{\rm pole}$ (right),
both at N$^2$LL  and NLL. The \Pythia datasets use $m_t^{\rm MC} = 173$\,GeV as an input. }
\end{center}
\end{figure}

\section{ Numerical Results of the Calibration}
To visualize the stability of our fits we display in Fig.~\ref{fig:histograms} the distribution of best-fit mass values obtained for 500
random profile functions for  \mbox{$m_t^{\rm MC} = 173$\,GeV} based on the $Q$ set $(600-1400)$ and the bin range $(60\%, 80\%)$. Results are
shown for $m_t^{\rm MSR}(1\,\mbox{GeV})$ and $m_t^{\rm pole}$ at NLL and N$^2$LL order, exhibiting good convergence, with the higher order
results having a smaller perturbative scale uncertainty. The results for $m_t^{\rm MSR}(1\,\mbox{GeV})$ are stable and about $200$\,MeV
below $m_t^{\rm MC}$ confirming the close relation of $m_t^{\rm MSR}(1\,\mbox{GeV})$ and $m_t^{\rm MC}$ suggested in
Refs.~\cite{Hoang:2008xm,Hoang:2014oea}. We observe that $m_t^{\rm pole}$ is about $1.1$\,GeV (NLL) and $0.7$\,GeV (N$^2$LL) lower than $m_t^{\rm MC}$, demonstrating that corrections here are bigger, and {\em that the MC mass cannot simply be identified with the pole mass}.
The results from the fits to the 21 different Q sets and bin ranges mentioned above are quite similar. Their differences can be interpreted as a quantification of
the level of incompatibility between the MC event generator results and the QCD predictions. Unlike the perturbative uncertainties they need not necessarily to decrease when going from NLL to N$^2$LL due to intrinsic limitations of the MC generator. We therefore use the differences from the 21 fits to assign an additional
{\it incompatibility uncertainty} between QCD and the MC generator for the calibration which one may interpret as an estimate for the intrinsic MC uncertainty. 

To quote final results we used the following procedure:
(1) Take the average of the highest and lowest central values from the 21 sets as the final central value of our calibration.
(2) Take the average of the scale uncertainties of these sets as our final estimate for the perturbative uncertainty.
(3) Take the half of the difference of the largest and smallest central values from the sets as the incompatibility uncertainty between
QCD and the MC.
\mbox{(4) Quadratically} add the perturbative, and incompatibility errors to obtain a final uncertainty.

Using $\alpha_s$ values within the uncertainty of the world average  $\alpha_s(m_Z)=0.1181(13)$ gives an additional parametric uncertainty
of $\simeq 20$\,MeV for $m_t^{\rm MSR}(1\,\mbox{GeV})$ and $m_t^{\rm pole}$ at N$ ^2$LL order. This is an order of magnitude smaller than
the other uncertainties and we therefore neglect it.
Table~\ref{tab:results} shows our final results for the MSR mass $m_t^{\rm MSR}(1\,\mbox{GeV})$ and $m_t^{\rm pole}$ at NLL and N$^2$LL order,
utilizing the $m_t^{\rm MC}=173$\,GeV dataset. For $m_t^{\rm MSR}(1\,\mbox{GeV})$ we observe a reduction of perturbative uncertainties from
$260$\,MeV at NLL to  $190$\,MeV at N$^2$LL. The corresponding incompatibility uncertainties are $140$ and $110$\,MeV. The corresponding fit
results for the first shape function moment are $\Omega_1^{\rm PY}=0.42\pm 0.07\pm0.03$\,GeV at N$^2$LL and
$\Omega_1^{\rm PY}=0.41\pm 0.07\pm0.02$\,GeV at NLL order with the first uncertainty coming from scale variation and second from 
incompatibility. The result agrees nicely with the expectation that $\Omega_1\sim \Lambda_{\rm QCD}$.

Using the pole mass scheme in our theory prediction and repeating the calibration fit we find that there is a significant difference to $m_t^{\rm MC}$, and we observe that the central value shifts by $330$\,MeV between NLL and N$^2$LL order, see Tab.~\ref{tab:results}. There is a reduction of perturbative uncertainties like in the MSR scheme, however the 
incompatibility uncertainty increases at N$^2$LL order. Interestingly, while we observe that $m_t^{\rm pole} < m_{t,1\,\rm GeV}^{MSR}$ for our fit result in the pole mass scheme, one obtains $m_t^{\rm pole} > m_{t,1\,\rm GeV}^{MSR}$ when converting our MSR mass result to the pole mass. At NLO the corresponding correction reads $m_t^{\rm pole} = m_{t,1\,\rm GeV}^{MSR} + 0.17\,{\rm GeV} + \mathcal{O}(\alpha_s^2)$. 
This observation may not be unexpected, since the pole mass often leads to poorer convergence of perturbative series due to the pole mass renormalon problem. This indicates that the uncertainty in the relation between $m_t^{\rm MC}$ and $m_t^{\rm pole}$ is actually larger than indicated by the pole mass results in Tab.~\ref{tab:results} by themselves.

\begin{table}[t!]
\begin{center}
\begin{tabular}{|llcccc|}
\multicolumn{6}{c}{\footnotesize $m_t^{\rm MC} = 173$\,GeV\ \ 
  \big($\tau_2^{e^+e^-}$\big)} \\
\hline
~mass~~~~~&order~~~& central & perturb.  &  incompatibility & ~total~\\
\hline\hline
~$m_{t,1\,\rm GeV}^{\rm MSR}$ & NLL      & $172.80$ & $0.26$ & $0.14$ & $0.29$\\
~$m_{t,1\,\rm GeV}^{\rm MSR}$ & N$^2$LL  & $172.82$ & $0.19$ & $0.11$ & $0.22$\\
~$m_t^{\rm pole}$             & NLL      & $172.10$ & $0.34$ & $0.16$ & $0.38$\\
~$m_t^{\rm pole}$             & N$^2$LL  & $172.43$ & $0.18$ & $0.22$ & $0.28$\\
\hline
\end{tabular}
\caption{\label{tab:results} Results of the calibration for $m_t^{\rm MC} = 173$\,GeV in \Pythia, combining results from all Q sets and bin
ranges. Shown are central values, perturbative and incompatibility uncertainties, and the total uncertainty, all in GeV. The results shown for the pole mass $m_t^{\rm pole}$ are not final as explained in the text.}
\end{center}
\end{table}

We have carried out the calibration procedure for $m_t^{\rm MC}$ values between $170$ and $175$\,GeV in steps of $1$\,GeV, and the outcome of our fits showed a behavior consistent with the results given in Tab.~\ref{tab:results}. In future studies such calibration results should be independently determined for different MC event generators and also for generator settings (such as different tunes). In the case of the MSR mass it is clear that the results from Tab.~\ref{tab:results} are well behaved perturbatively and we therefore expect that higher order results will lie within the given uncertainties. This means that for the setup described before a MC top quark mass value of \mbox{$m_t^{\rm MC} = 173$\,GeV} (as a representative example) can be associated with the MSR mass value of $m_{t,1\,\rm GeV}^{\rm MSR} = 172.82\pm0.22$\,GeV. Overall we find that
\begin{equation}
m_t^{\rm MC} = m_{t,1\,{\rm GeV}}^{\rm MSR} + (0.18 \pm 0.22)\,{\rm GeV}\,,
\end{equation}
for MC top quark masses between $172$ and $174\,{\rm GeV}$. So $m_t^{\rm MC}$ and $m_{t,1\,{\rm GeV}}^{\rm MSR}$ may be identified to within about $200\,{\rm MeV}$.

To give a definite result including a reliable uncertainty estimate for the relation of $m_t^{\rm MC}$ and the top quark pole mass (that accounts for the observations discussed above) we adopt the following strategy at NNLL: (1)~We use the perturbative MSR-pole mass relation at NLO quoted above and take the average between the direct pole mass calibration result from Tab.~\ref{tab:results} (lowest line) and the converted result (using the MSR mass result as input) for the final pole mass central value. (2)~For the uncertainty we take the quadratic sum of the direct determination uncertainty shown in Tab.~\ref{tab:results} (lowest line) and half of the difference with respect to the MSR converted result (which amounts to $0.28\,{\rm GeV}$ at NNLL).

Applying this procedure we find that for the above described setup a MC top quark mass value of \mbox{$m_t^{\rm MC} = 173$\,GeV} can be associated with the pole mass value of \mbox{$m_{t}^{\rm pole,\rm N^2LL} = 172.71\pm0.40$\,GeV.} To obtain the relation of $m_t^{\rm MC}$ and $m_t^{\rm pole}$ at NLL order we proceed in the analogous way by using the direct calibration result for the pole mass in Tab.~\ref{tab:results} (second-to-lowest line) and the tree-level relation $m_t^{\rm pole} = m_{t,1\,\rm GeV}^{\rm MSR}$. Here, half of the difference of the two resulting values for $m_t^{\rm pole}$ amounts to $0.35$\,GeV. For $m_t^{\rm MC} = 173$\,GeV the pole mass NLL result reads $m_t^{\rm pole,NLL} = 172.45\pm0.52$\,GeV, which is consistent with the NNLL result. Overall we find that $m_t^{\rm MC} = m_t^{\rm pole} + (0.29 \pm 0.40)\,{\rm GeV}$ at NNLL order for MC top quark masses between $172$ and $174\,{\rm GeV}$. So $m_t^{\rm pole}$ and $m_{t,1\,{\rm GeV}}^{\rm MSR}$ may also be identified - however, only within about $400\,{\rm MeV}$.

To the extent that the treatment of the top in MC generators and QCD factorizes for different kinematically sensitive observables and from whether one considers $e^+e^-$ or $pp$ collisions, our method can be used to calibrate $m_t^{\rm MC}$ in current experimental reconstruction analyses. It is nevertheless important to clarify to which extent the MC top quark mass determined for one process and observable has this universal meaning. A promising approach to this is to do a similar analysis as the one presented in this work while considering a different process e.g. $pp$ or a sufficiently different $e^+e^-$ observable. $pp$ collisions introduce initial state radiation, color reconnection, and additional hadronization and multi-parton interaction effects, not present in $e^+e^-$.  A possible way around the most troublesome complications for $pp$ is to use so called ``light jet-grooming'' as in Ref.~\cite{Hoang:2017kmk}. Alternatively, it is also interesting to explore sufficiently different observables for $e^+e^-$ which will be described in the following. Prior to this, we believe that applying our $e^+e^-$ calibration to $m_t^{\rm MC}$ from a typical $pp$ reconstruction analysis will give a more accurate result than simply assuming $m_t^{\rm MC} = m_t^{\rm pole}$.

\subsection*{Exploring MC top quark mass universality with different $e^+e^-$ observables}

One possibility for an alternative $e^+e^-$ observable is a mass-sensitive version of C-parameter, defined as:
\begin{equation}
   C_{\rm M} = \frac{3}{2}\left[2-\frac{1}{Q^2}\sum_{i\neq j}\frac{(p_i\cdot p_j)^2}{p_i^0 p_j^0}\right] \,.
\end{equation}

The $C_M$ observable coincides with the regular C-parameter definition for massless final state particles and was already used in Ref.~\cite{Gardi:2003iv} due to its favorable features when looking at massive final state particles. The biggest difference compared with thrust is the strong sensitivity to the top quark decay. Due to the clustering property of thrust, the events where the decay products respect the original hemisphere division (which give the leading contribution) contribute in the same way to the thrust distribution as in the case of a stable top. This is why the decay inclusive formula from Eq.~\eqref{eq:factheo} works well in this case. For the mass-sensitive C-parameter (subsequently we will refer to this simply as C-parameter) a more differential treatment of the decay is needed to describe the unstable distribution correctly. This fact becomes obvious when looking at the difference in peak position in the stable $C_{\rm M}^{\rm min} = 12(m_t/Q)^2[1-(m_t/Q)^2]$ and unstable case \mbox{$C_{\rm M}^{\rm min} \approx 12(m_t/Q)^2[1- 4(m_t/Q)^2]$}.

For the leading contribution, which comes from those events where the decays respect the hemisphere division, the factorization theorem for the singular part can be extended to the following form:
\begin{align}\label{eq:cunstable}
 \frac{1}{\sigma_0}\frac{{\rm d}\sigma^{\rm unstable}}{{\rm d} C_M} &= \int{\rm d}\bar{C}\frac{1}{\sigma_0}\frac{{\rm d}\sigma^{\rm stable}_{\rm s}}{{\rm d} C_M}(C_{\rm M} - \bar{C}) F(\bar{C})
 \,,
\end{align}
where $\frac{1}{\sigma_0}\frac{{\rm d}\sigma^{\rm stable}}{{\rm d} C_M}$ represents the C-parameter factorization formula for stable top quarks in analogy to Eq.~\eqref{eq:factheo}. As indicated in Eq.~\eqref{eq:cunstable} this expression needs to be convoluted with an additional function $F$, called the \emph{decay function}. It is given by \mbox{$F(\bar{C}) = \int \prod_{i}{\rm d}\phi_i{\rm d}\theta_i \,\delta(\bar{C} - C_M^{\rm decay}(\theta_i,\phi_i) + C_{\rm M}^{\rm min})$} with $C_M^{\rm decay}(\theta_i,\phi_i)$ giving the value of the C-parameter for the (anti-)top decay products and $\phi_i$ and $\theta_i$ denoting the polar and azimuthal angles of the decay products in the rest frame of the decaying particle. This function may account for NLO QCD corrections to the top quark decay and basically accounts for the C-parameter difference between the stable case and the unstable case where the (anti-)top quark is produced at LO as an on-shell particle. At LO it can be obtained from standard fixed-order Monte Carlo programs. This approach is exact in the limit that top production and decay factorize and allows to treat the radiation off the top with analytical tools, as in the case of thrust and subsequently takes care of the (anti-)top quark decay in a more differential manner. With this setup a comparably precise calibration should be feasible while considering a more decay sensitive, hence significantly different observable and will be discussed in more detail in an upcoming publication~\cite{Hoang:2018}.

\subsection*{Acknowledgment}

We acknowledge partial support by the FWF Austrian Science Fund under the Doctoral Program No.~W1252-N27 and the Project No.~P28535-N27, the Spanish MINECO ``Ram\'on y Cajal'' program
(RYC-2014-16022), the U.S. Department of Energy under the Grant No.~DE-SC0011090, and the Simons Foundation through the Grant 327942. We also thank the Erwin-Schr\"odinger International Institute for Mathematics and Physics, the University of Vienna and Cultural Section of the
City of Vienna (MA7) for partial support.

\bibliographystyle{JHEP}
\bibliography{../thrust3}

\end{document}